\documentclass [showpacs,amsmath,amssymb,12pt]{revtex4}
\usepackage{graphics}
\usepackage{dcolumn}
\usepackage{bm}
\usepackage{textcomp}
\begin {document}

\title {\large{Eigenvalue problem in two dimension for an irregular boundary}}
\author{S. Chakraborty}
\email{somdeb.chakraborty@saha.ac.in}
\affiliation {Saha Institute of Nuclear Physics, 1/AF,
BidhanNagar, Kolkata 700064, India}

\author{ J. K. Bhattacharjee}
\email{jayanta.bhattacharjee@gmail.com}
\affiliation{ S.N. Bose National Centre for Basic Sciences,
Sector-III, Block-JD, Salt Lake, Kolkata 700098, India}

\author{S. P. Khastgir}
\email{pratik@phy.iitkgp.ernet.in}
\affiliation{ Department of Physics and Meteorology,
Indian Institute of Technology, Kharagpur 721302, India}

\begin {abstract}
An analytical perturbative method is suggested for determining
the eigenvalues of the Helmholtz equation $(\nabla^{2}+k^{2})\psi=0$
in two dimensions where $\psi$ vanishes on an irregular closed curve.
We can thus find the energy levels of a
quantum mechanical particle confined in an infinitely deep
potential well in two dimensions having an irregular boundary
or the vibration frequencies of a membrane whose edge is an
irregular closed curve. The method is tested by calculating the
 energy levels for an elliptical and a supercircular boundary
and comparing with the results obtained numerically. Further, the
phenomenon of level crossing due to shape variation is also discussed.
\end {abstract}
\maketitle
\newpage
 \section{Introduction}
 The energy levels of a
quantum particle confined in a 2D regular box can be solved
exactly only in the cases of a square and a triangle and in the
limiting case of a circle. While the determination of the energy
levels for the circular or the square boundary is a trivial
exercise, the problem of the triangular boundary is more
formidable \cite{Krishnamurthy}. The corresponding problems in the
classical regime can be the flow of liquid through a pipe of
polygonal cross-section or the free vibration of a membrane (with
a fixed boundary) of polygonal shape. The classical problems, like
their quantum counterparts, are amenable to simple analytical
treatments only in the cases of a circle, a square and a triangle.
The problem of a regular polygonal box has been solved by
perturbing about the equivalent circle and the results have been
quite accurate \cite{Jayanta}. The same problem has been solved by
Cureton and Kuttler \cite{Cureton} in the context of vibration of
membranes. Here we address the problem of finding out the energy
eigenvalues when the boundary has no simple geometric shape. The
Schr\"odinger equation for a particle of mass $m$ and energy $E$
confined in an infinitely deep 2D potential well is,
\begin{subequations}
\begin {equation}
-{{\hbar^2}\over{2m}}\nabla^{2}\psi= E\psi. \label{sh}
\end {equation}
\noindent The above equation can be recast as,
\begin {equation}
(\nabla^{2}+k^{2})\psi=0,
\end {equation}
\end{subequations}
where $k=\sqrt{\frac{2mE}{\hbar^2}}$. Thus the problem boils down
to solving the Helmholtz equation with the Dirichlet condition
$\psi=0$ on the `irregular' boundary. Exact solutions can be
obtained only in a few special cases as mentioned earlier. The
standard procedure is to choose a curvilinear coordinate system
suited to the geometry of the problem and employ the method of
separation of variables. For a boundary having an irregular shape
no particular coordinate system will be useful. Hence, we  resort
to perturbative methods to solve the problem. Here we will perturb
the boundary about a circle so that in our problem solutions can
be obtained in the form of corrections to the solutions for the
circular boundary. Till now, most of the efforts at finding out
the eigenvalues of the Helmholtz equation for an irregular
boundary have been numerical. Mazumdar \cite{M1, M2, M3} reviews
the approximate methods invoked for this problem. In addition to
the extensive summary of theoretical results, Kuttler and
Sigillito \cite{Kuttler} also give a comprehensive review of the
different numerical methods employed. More recently, Amore
\cite{Amore} gives a numerical recipe using a collocation approach
based on little sinc functions. As far as analytical works are
concerned, Rayleigh \cite{Rayleigh} and also Fetter and Walecka
\cite{Fetter} find the ground state energy eigenvalues for a
vibrating membrane. A general formalism has been suggested by
Morse and Feshbach \cite{Morse} using the Green functions. Parker
and Mote \cite{Parker} have put forward a perturbative method for
finding the eigenvalues and the eigenfunctions through fifth
order. A similar method has been proposed by Nayfeh \cite{Nayfeh}.
However, the eigenvalues are found out only to the first order.
Read \cite{Read} has also suggested a general analytical approach
to the problem. Bera et al \cite{Bera} have proposed a
perturbative approach to the problem  but  failed to express the
solutions in a closed form. Our approach is similar in spirit to
that of Bera. Here we present a  solution to the problem in a more
systematic and efficient manner. The perturbative correction to
the eigenvalues and the eigenfunctions are presented in a closed
form at each order of perturbation. The method is tested by
comparing the analytical results with those obtained numerically
for a supercircular and an elliptical boundary. Further, the
phenomenon of energy level crossing as induced by the shape
variation is also dealt with for both the boundaries. In section
II we set up our general scheme and in section III we apply it to
the cases of a supercircle and an ellipse. A short conclusion is
presented in section IV.

\section{Perturbation about the Equivalent Circle}
It was shown by Rayleigh \cite{Rayleigh} that the fundamental
frequency of a membrane whose boundary is not extravagantly
elongated is nearly same as that of a mechanically similar
circular membrane having the same area. The above result naturally
leads us to develop, in following, a perturbation about the equal
area circle.
Given, any $r(\theta)=r(\theta+2\pi)$, defining the boundary in 2D
enclosing an area, $A={1\over 2}\int_0^{2\pi}r^2(\theta)d\theta$,
we first construct a circle of radius, $R_{0}$, such that,
\begin{equation}
A=\pi R_{0}^{2}\label{circ}.
\end{equation}
We can then expand $r(\theta)$ about $R_{0}$ in terms of Fourier
series at different orders of smallness (denoted by $\lambda$) as,
\begin{equation}
r(\theta)=R_0\left[1+\sum_{\sigma=1}^{\infty}
\lambda^{\sigma}f^{(\sigma)}(\theta)\right],
\label{rex}
\end{equation}
where,
\begin{equation}
f^{(\sigma)}(\theta)=\sum_{n=0} ^{\infty}(C_{n}^{(\sigma)}\cos
n\theta +S_{n}^{(\sigma)} \sin n\theta).  \label{gfs}
\end{equation}
Here, for simplicity, we have considered a one parameter
(deformation parameter), $\lambda$, dependence of $r(\theta)$,
which thus represents a family of curves which reduce to the
equation for a circle in the limiting case $ \lambda \to 0$. In
principle, $\lambda$ should be much smaller than unity ensuring
that the variation of $r(\theta)$ with $\theta$ is small enough to
permit the use of perturbative methods. However, as we will see in
the next section that for the case of a supercircle $\lambda\sim
1$ works quite well and keeps the results within 10\% error. We
also note here that the Fourier expansion of the boundary in
(\ref{rex}) is rather unusual and which makes our method different
from all other existing methods. Here in fact each
$f^{(\sigma)}(\theta)$ is a Fourier series in itself of
order $\lambda^\sigma$. Earlier methods in the literature had
worked with only one Fourier series - that is by summing all the
orders into one. The main advantage in treating the problem like
this is to have an analogy with the time independent perturbation
scheme of quantum mechanics and obtain closed form solutions at
each order of $\lambda$. If we now calculate the area using
(\ref{rex}), (\ref{gfs}) and equate it with $\pi R_0^2$, we arrive
at the following constraint relations
 among the Fourier coefficients,
\begin{subequations}
\begin{equation}
\sum_{n=0}^{\infty}\sum_{\nu=1}^{\sigma-1}\left[C_{n}^{(\nu)}
C_{n}^{(\sigma-\nu)}+2C_{0}^{(\nu)}C_{0}^{(\sigma-\nu)}+
S_{n}^{(\nu)}S_{n}^{(\sigma-\nu)}\right] = -4C_{0}^{(\sigma)}.
\end{equation}
In particular we have,
\begin{equation}
C_{0}^{(1)}=0,
\end{equation}
and
\begin{equation}
4C_{0}^{(2)}=-\sum_{n=1}^{\infty}[C_{n}^{(1)2}+S_{n}^{(1)2}].\label{co2}
\end{equation}
\end{subequations}
Now, as a first approximation, the energy $E_{0}$ of the particle
confined by $r(\theta)$ will be that of a particle enclosed in a
circle of radius $R_{0}$,
\begin{equation}
E_{0}=\frac{\hbar^{2}\rho_{l,j}^{2}}{2mR_{0}^{2}},    \label{gsec}
\end{equation}
with $\rho_{l,j}=k_{l,j}R_{0}$ being the $j^{\rm th}$ node of the
$l^{\rm th}$ order Bessel function. The next step  is to improve
upon the `equal area' approximation by perturbing the equivalent
circle and finding out the first and the second order corrections
to the
eigenvalues.\\
We now treat $\lambda$ as the perturbation parameter and expand
$\psi$ and $E$ as,
\begin{subequations}
\begin{equation}
\psi=\psi_{0}+\lambda\psi_{1}+\lambda^{2}\psi_{2}+..., \label{ex1}
\end{equation}
\begin{equation}
E=E_{0}+\lambda E_{1}+\lambda^{2}E_{2}+... ~. \label{ex2}
\end{equation}
\end{subequations}
Using (\ref{ex1}), (\ref{ex2}) in (\ref{sh}), equating the
coefficients of different powers of $\lambda$ to 0 and after some
rearrangement we arrive at the set of equations,
\begin{subequations}
\begin{equation}
(\nabla^{2} +\frac{2mE_{0}}{\hbar^{2}})\psi_{0}=0, \label{0}
\end{equation}
\begin{equation}
(\nabla^{2}+\frac{2mE_{0}}{\hbar^{2}})\psi_{1}
=-\frac{2mE_{1}}{\hbar^{2}}\psi_{0}, \label{1}
\end{equation}
\begin{equation}
(\nabla^{2}+\frac{2mE_{0}}{\hbar^{2}}
)\psi_{2}=-\frac{2m}{\hbar^{2}}({E_1}{\psi_1}+{E_2}{\psi_0}).
\label{2}
\end{equation}
\end{subequations}
Equation (\ref{0}) can readily be identified as the equation for
the circular boundary with $ \psi_{0}$ as
the eigenfunction corresponding to energy $E_{0}$.\\
The boundary condition is,
 $$\psi(R_{0}+\lambda R_{0}f^{(1)}+
\lambda^{2}R_{0}f^{(2)}+...)=0.$$\\
Taylor expanding about $r=R_{0}$, with (\ref{ex1}) and equating
the coefficients of different powers of $\lambda$ to 0, we find,
\begin{subequations}
\begin{equation}
\psi_{0}(R_{0})=0,  \label{bc0}
\end{equation}
\begin{equation}
\psi_{1}(R_{0})+R_{0}f^{(1)}\psi_{0}^{'}(R_{0})=0,   \label{bc1}
\end{equation}
\begin{equation}
\psi_{2}(R_{0})+R_{0}f^{(1)}\psi_{1}^{'}(R_{0})+R_{0}f^{(2)}\psi_{0}^{'}(R_{0})+
\frac{1}{2}R_{0}^{2}f^{(1)2}\psi_{0}^{''}({R_0})=0.    \label{bc2}
\end{equation}
\end{subequations}
We discuss separately the cases $l=0$ and $l \neq 0$.

\subsection{Calculation of  Energy for $l=0$ State}
For the $l=0$ state,
\begin{equation}
\psi_{0}=NJ_{0}(\rho), \label{sol0}
\end{equation}
where $\rho=kr$, $J_{0}$ is the $0^{\rm th}$ order Bessel
function, and $N=1/({\sqrt{\pi}R_0J_1(\rho_{0,j})})$, is the
normalisation constant. $E_{0}$ is obtained from (\ref{gsec}) with
$l=0$, and an appropriate $j$, as $\psi_0$ satisfies boundary
condition (\ref{bc0}). The first order correction to the wave
function, obtained as a solution to (\ref{1}) is,
\begin{equation}
\psi_{1}=\sum_{p=1}^{\infty}(a_{p}\cos p\theta+\bar a_{p}\sin
p\theta )J_{p} +a_{0}J_{0}-\frac{\rho E_{1}}{2E_{0}}NJ_{1},
\label{gfc}
\end{equation}
where the last term is the particular integral to (\ref{1}).
Incorporating (\ref{gfc}) in (\ref{bc1}) and separately matching
the coefficients of the cosine and the sine terms we have,
\begin{subequations}
\begin{equation}
a_{p}=-\rho_{0,j}NC_{p}^{(1)}\frac{J_{0}'(\rho_{0,j})}{J_{p}(\rho_{0,j})},
\label{gfcc}
\end{equation}
\begin{equation}
\bar a_{p}=-\rho_{0,j}NS_{p}^{(1)}\frac{J_{0}'(\rho_{0,j})}{J_{p}(\rho_{0,j})},
\label{gfsc}
\end{equation}
\begin{equation}
E_{1}=0.  \label{gse1c}
\end{equation}
\end{subequations}
The remaining constant $a_{0}$ can be found out by normalising the
corrected wave function over the enclosed area. However, that is
not required right now for our purpose. (12c) implies that there
cannot be any correction to the energy in the first order. So any
possible correction to the energy can only come from the second or
higher orders.
In a similar fashion the second correction to the wave function as
a solution to (\ref{2}) with $E_{1}=0$ is found out to be,
\begin{equation}
\psi_{2}=\sum_{p=1}^{\infty}(b_{p}\cos p\theta+\bar b_{p}\sin
p\theta )J_{p} +b_{0}J_{0}-\frac{\rho E_{2}}{2E_{0}}NJ_{1},
\label{gsc}
\end{equation}
which, when introduced in (\ref{bc2}), now yields,
\begin{subequations}
\begin{equation}
E_{2}=E_{0}\left[\sum_{k=1}^{\infty}(C_{k}^{(1)2}+
S_{k}^{(1)2})\left[\frac{1}{2}
+\rho_{0,j}\frac{J_{k}'(\rho_{0,j})}{J_{k}(\rho_{0,j})}
\right]-2C_{0}^{(2)}\right],       \label{gse2c}
\end{equation}

$$b_{p}=-\rho_{0,j}\frac{J_{0}'(\rho_{0,j})}{J_{0}(\rho_{0,j})}\left[NC^{(2)}_{p}+a_{0}C_{p}^{(1)}\right]$$
$$+
\frac{\rho_{0,j}J_{0}'(\rho_{0,j})N}{2J_{0}(\rho_{0,j})  }\sum_{k=1}^\infty[C_{p+k}^{(1)}
C_{k}^{(1)}+S_{p+k}^{(1)}S_{k}^{(1)} +C_{|p-k|}^{(1)}
C_{k}^{(1)}$$
\begin{equation}
- S_{p-k}^{(1)}S_{k}^{(1)}+S_{k-p}^{(1)}S_{k}^{(1)}]
\left(\frac{1}{2}+\rho_{0,j}\frac{J_{k}'(\rho_{0,j})}{J_{k}(\rho_{0,j})}\right),
\end{equation}

$$\bar b_{p}=-\rho_{0,j}\frac{J_{0}'(\rho_{0,j})}{J_{0}(\rho_{0,j})}\left[NS_{p}^{(2)}+a_{0}S_{p}^{(1)}\right]$$
$$+
\frac{\rho_{0,j}J_{0}'(\rho_{0,j})N}{2J_{0}(\rho_{0,j})}\sum_{k=1}^\infty[S_{p+k}^{(1)}
C_{k}^{(1)}-C_{p+k}^{(1)}S_{k}^{(1)} +C_{|p-k|}^{(1)} S_{k}^{(1)}
$$
\begin{equation}
- S_{k-p}^{(1)} C_{k}^{(1)}+S_{p-k}^{(1)}C_{k}^{(1)}]
\left(\frac{1}{2}+\rho_{0,j}\frac{J_{k}'(\rho_{0,j})}{J_{k}(\rho_{0,j})}\right).
\end{equation}
\end{subequations}
As before, the remaining constant $b_{0}$ can be determined by
normalising the wave function up to the  order of $\lambda^{2}$.
\subsection{Calculation of  Energy for $l \neq 0$ state}
The $l \neq 0$ states come in 2 varieties,
\begin{equation}
\psi_{0}=N_{l}J_{l}(\rho)\left(\begin{array}{c} \cos l\theta \\
\sin l\theta \end{array} \right),
\end{equation}
where, $N_l=\sqrt{2}/({\sqrt{\pi}R_0J'_l(\rho_{l,j})})$. $E_0$ is
given by (\ref{gsec}). For simplicity, we assume that
$S_{n}^{(\sigma)}=0$ for all $\sigma$. We shall first work with
\begin{equation}
\psi_{0}=N_{l}J_{l}(\rho) \cos l\theta.
\end{equation}
The result for the other case will be similar. The first
correction to the wave function obtained as a solution to
(\ref{1}) is,
\begin{equation}
\psi_{1}=\sum_{p=0,p\neq l}^{\infty}a_{p}J_{p}\cos p\theta+
\left(a_{l}J_{l}-\frac{E_{1}}{E_{0}}\frac{\rho}{2}N_{l}J_{l+1}\right
)\cos l\theta.
\end{equation}
Following a similar procedure as that for the ground state we now
have,
\begin{subequations}
\begin{equation}
E_{1}=-C_{2l}^{(1)}E_{0},  \label{ese1cc}
\end{equation}
\begin{equation}
a_{p}=-\frac{\rho_{l,j}}{2}N_{l}\frac{J_{l}'(\rho_{l,j})}{
J_{p}(\rho_{l,j})}  (C_{p+l}^{(1)}+ C_{|p-l|}^{(1)}),   ~~~~ \text
{for $p\neq 0,l$},
\end{equation}
\begin{equation}
a_{0}=-\frac{\rho_{l,j}}{2}N_{l}\frac{J_{l}'(\rho_{l,j})}{
J_{0}(\rho_{l,j})}  C_{l}^{(1)}.
\end{equation}
\end{subequations}
$a_{l}$ can be obtained from the normalisation condition. The
second order corrections yield,
\begin{subequations}
\begin{equation}
  \psi_{2}=\sum_{m=0}^{\infty}\left[b_{m}J_{m}-\frac{\rho E_{1}}
{2E_{0}}a_{m}J_{m+1}\right]\cos m\theta 
+\left[C_{2l}^{(1)2}\frac{\rho}{4}J_{l+2}
-\frac{E_{2}}{E_{0}}J_{l+1}\right]N_{l}\frac{\rho}{2} \cos
l\theta,
\end{equation}
$$\frac{E_{2}}{E_{0}}=\frac{C_{2l}^{(1)2}}{2}+\frac{1}{4}
\sum_{n=1}^\infty
C_{n}^{(1)}(2C_{n}^{(1)}+C_{2l+n}^{(1)}+C_{|2l-n|}^{(1)})
-2C_{0}^{(2)} $$
\begin{equation}
 -C_{2l}^{(2)} +\sum_{{n=1}\atop{n\neq l}}^{\infty}(C_{n+l}^{(1)}
 +C_{|n-l|}^{(1)})^{2}\frac{\rho_{l,j}J_{n}'(\rho_{l,j})}{2J_{n}(\rho_{l,j})}+
C_{l}^{(1)2}\frac{\rho_{l,j}J_{0}'(\rho_{l,j})}{J_{0}(\rho_{l,j})}.\label{ese2cc}
\end{equation}
\end{subequations}
The constants $b_{m}$ can also be determined as in the case of the
ground state. However, they are not needed for now. Need for them
would arise when one would evaluate the third order correction for
energy. For the case,
\begin{equation}
\psi_{0}=N_{l}J_{l}(\rho)\sin l\theta,
\end{equation}
similar calculations result in,
\begin{subequations}
\begin{equation}
\psi_{1}=\sum_{p=1,p\neq l}^{\infty}\bar a_{p}J_{p}\sin p\theta+
\left(\bar a_{l}J_{l}-\frac{E_{1}}{E_{0}}
\frac{\rho}{2}N_{l}J_{l+1}\right )\sin l\theta,
\end{equation}
\begin{equation}
E_{1}=C_{2l}^{(1)}E_{0}, \label{ese1cs}
\end{equation}
\begin{equation}
\bar a_{p}=\frac{\rho_{l,j}}{2}N_{l}\frac{J_{l}'(\rho_{l,j})}{
J_{p}(\rho_{l,j})}  (C_{p+l}^{(1)}- C_{|p-l|}^{(1)}),   ~~~~ \text
{for $p\neq l$},
\end{equation}
\begin{equation}
  \psi_{2}=\sum_{m=1}^\infty\left[\bar b_{m}J_{m}
-\frac{\rho E_{1}}{2E_{0}}\bar a_{m}J_{m+1}\right]\sin m\theta 
+\left[C_{2l}^{(1)2}\frac{\rho}{4}J_{l+2}-\frac{E_{2}}{E_{0}}
J_{l+1}\right]N_{l}\frac{\rho}{2} \sin l\theta,
\end{equation}
and
$$  \frac{E_{2}}{E_{0}}=\frac{C_{2l}^{(1)2}}{2}+\frac{1}{4}
\sum_{n=1}^\infty
C_{n}^{(1)}(2C_{n}^{(1)}-C_{2l+n}^{(1)}-C_{|2l-n|}^{(1)})
-2C_{0}^{(2)} $$
\begin{equation}
 +C_{2l}^{(2)} +\sum_{n=1,n\neq l}^\infty \left(C_{n+l}^{(1)}
 -C_{|n-l|}^{(1)}\right)^{2}\frac{\rho_{l,j}
 J_{n}'(\rho_{l,j})}{2J_{n}(\rho_{l,j})}. \label{ese2cs}
\end{equation}
\end{subequations}
We do not give the expressions for $\bar b_{m}$, as they are not
needed now.

\begin{figure}[t]
\centering
{\scalebox{0.6}{\includegraphics{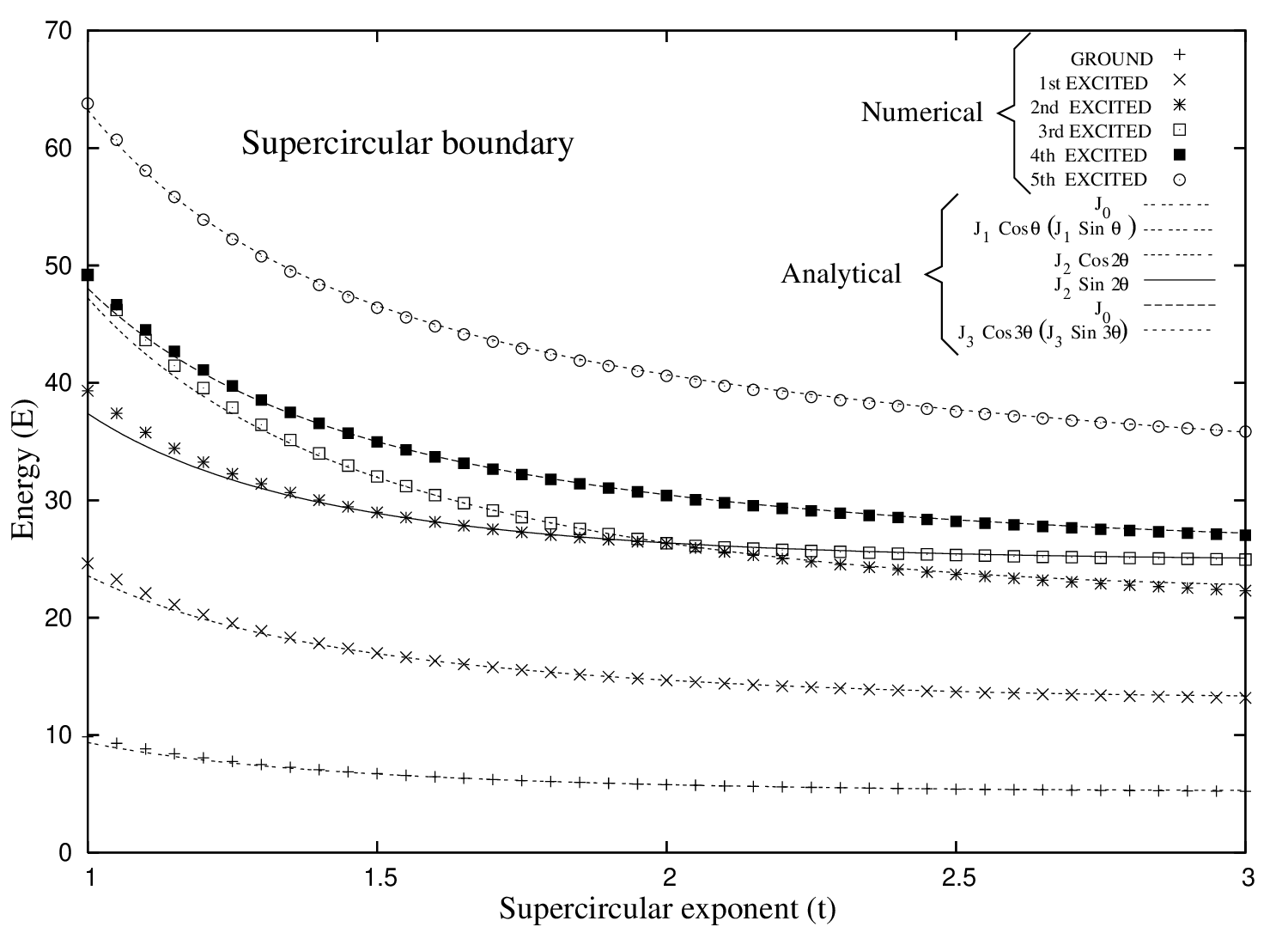}}}
\caption{\label{fig:2} Comparison of the energy eigenvalues
obtained numerically and analytically for a particle enclosed in a
supercircular boundary shown for the first 6 states (in units of
$\frac{\hbar^{2}}{2mR_{o}^{2}}$). Deformation parameter
$\lambda=2-t$.}
\end{figure}

\begin{figure}[t]
\centering
{\scalebox{1.0}{\includegraphics{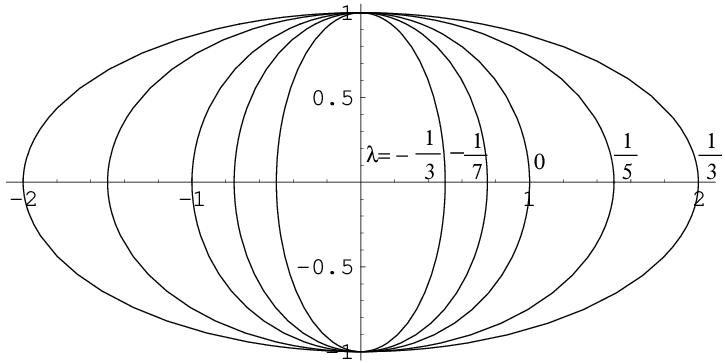}}}
\caption{\label{fig:3} Shape of the ellipse for different values
of $\lambda=\frac{a-b}{a+b}.$}
\end{figure}

{\section {Application to Simple Cases}}
The general formalism having been outlined above we now estimate
the energy levels of a supercircle  and an ellipse where direct
comparison with the numerical results can be made. Numerical
results were calculated using the finite difference method. Both
square and triangular grids were used separately for the numerical
simulation. The results agree quite well for both types of grids.
{\subsection{Particle Enclosed in a Supercircular Enclosure}} Piet
Hein Superellipse \cite{Gardner} is a special case of Lam\'e
curves described by,
\begin{equation}
\frac{|x|^{t}}{a^{t}}+ \frac{|y|^{t}}{b^{t}}=1,
\end{equation}
with $t>1$. $a$ and $b$ are positive real numbers. They are also
known as Lam\'e curves or Lam\'e ovals \cite{Gridgeman}.
Superellipses can be parametrically described as,
\begin{equation}
x=a\cos^{{2}/{t}}\phi,~~{\rm and}~~ y=b\sin^{{2}/{t}}\phi.
\end{equation}
Different values of $t$ would give us closed curves of different
shapes. For $t>1$ we consider only the real positive values of $
\cos^{{2}/{t}}\phi$ and $\sin^{{2}/{t}}\phi$ for $0 \leqslant \phi
\leqslant \frac{\pi}{2}$ and use the symmetry of the figure to
continue to the other quadrants. We are interested in the case
$a=b$, which corresponds to a supercircle. In polar coordinates
the equation for the supercircle is,
\begin{equation}
r=\frac{a}{(\cos^{t}\theta+\sin^{t}\theta)^{\frac{1}{t}}},\label{radius}
\end{equation}
and the radius of the equal area circle is,
\begin{equation}
R_{0}=a\sqrt{\frac{2}{t\pi}}\frac{[\Gamma(\frac{1}{t})]}
{\sqrt{[\Gamma(\frac{2}{t})]}}.
\end{equation}
\begin{figure}[t]
\centering 
{\scalebox{0.9}{\includegraphics{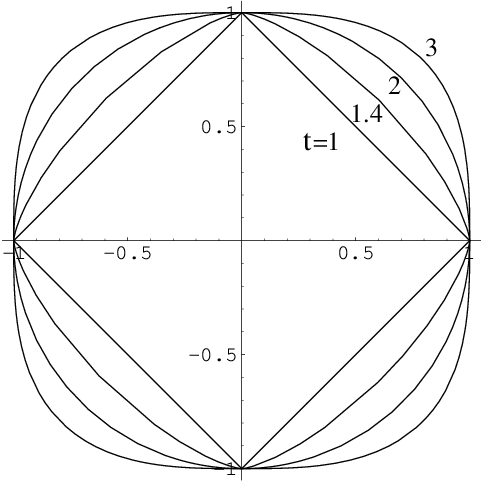}}}
\caption{\label{fig:1} Shape of the supercircle for different
values of $t$}.
\end{figure}
The shapes of supercircles for different values of $t$ are shown
in FIG.\ref{fig:1}. $t=2$ describes a circle of unit radius. In
this case we have a natural deformation parameter, $\lambda=2-t$.
Now $r(\theta)$, given by (\ref{radius}) can be Fourier expanded
and after some calculation one arrives  at the following,
\begin{equation}
r(\theta)=R_{0}[ 1+
\lambda\sum_{n=1}^{\infty}C_{4n}^{(1)}\cos 4n\theta
+ \lambda^{2} \sum_{n=0}^{\infty} C_{4n}^{(2)}\cos 4n\theta + {\rm
O}(\lambda^3)...],
\end{equation}
where the Fourier coefficients are found to be,
$$C_{4n}^{(1)}=-\frac{1}{4n(4n^2-1)},$$
$$C_{0}^{(2)}=-\frac{1}{4}\sum_{n=1}^{\infty}
 \left[\frac{1}{4n(4n^2-1)}\right]^{2}$$
 $$ \phantom{aaaa\qquad}={\frac{1}{16}}
\left(1-{\frac{5\pi^2}{48}}\right)=-0.0017552,$$
 using (\ref{co2}) and
 $$C_{4}^{(2)}=\frac{1}{32}
 \left( \frac{3\pi^{2}}{8}-\frac{23}{9}\right)=0.0357983.$$
Using these  Fourier coefficients, the first six energy levels are
calculated for the supercircular boundary in the range $-1
\leqslant \lambda \leqslant 1$, and compared with the numerically
obtained values. This is shown in FIG.\ref{fig:2}.
The numerical results are shown by discrete points and the analytical ones by
the continuous lines. The fact that even for such a wide range of
$\lambda$ the analytical results are in fairly good agreement with
those obtained numerically does indeed justify the validity of our
formalism. We see that for $|\lambda|$ as large as 1 the
deviations of analytical values from the numerical ones are within
10\%. Furthermore, it is to be noted that the energy level
corresponding to the unperturbed wave function $
\psi_{0}=N_{2}J_{2}\cos 2\theta$ is strongly affected compared to
the others and crosses over to its counterpart $
\psi_{0}=N_{2}J_{2}\sin 2\theta$ at $\lambda=0$. This crossing of
energy levels is solely induced by the variation in the shape
of the boundary of the potential well.\\
{\subsection {Particle Enclosed in an Elliptical Enclosure}} The
determination of the eigenvalues of the Helmholtz operator
 in 2D with an elliptical boundary has been investigated extensively.
 In this case the variable separation is possible
 in elliptical coordinate system and the
 problem is exactly solvable in principle.
 The problem reduces to solving the Mathieu
 differential equation for each of the separated coordinates.
 Extracting out the eigenvalues and the eigenfuctions from the above
 is a difficult task and often one relies on numerical estimation.
  So far most of the efforts have been directed at the numerical
 estimation of the eigenvalues \cite{Wilson, Hettich, Troesch}.
 Recently, an analytical method has been suggested by Wu and
 Shivakumar \cite{Yan}. Here we propose a simpler
 approach to the problem by our perturbative method.
The equation for an ellipse with semi-axes $a$ and $b$, in polar
coordinates is,
\begin{equation}
r(\theta)=\frac{b}{\sqrt{1-(1-\frac{b^2}{a^2})\cos^2\theta}},
\label{eqe1}
\end{equation}
Defining the deformation parameter,
\begin{equation}
\lambda=\frac{a-b}{a+b},
\end{equation}
we show the shapes of the ellipses for different values of
$\lambda$ in FIG.\ref{fig:3}.

\begin{figure}[t]
\centering
{\scalebox{0.6}{\includegraphics{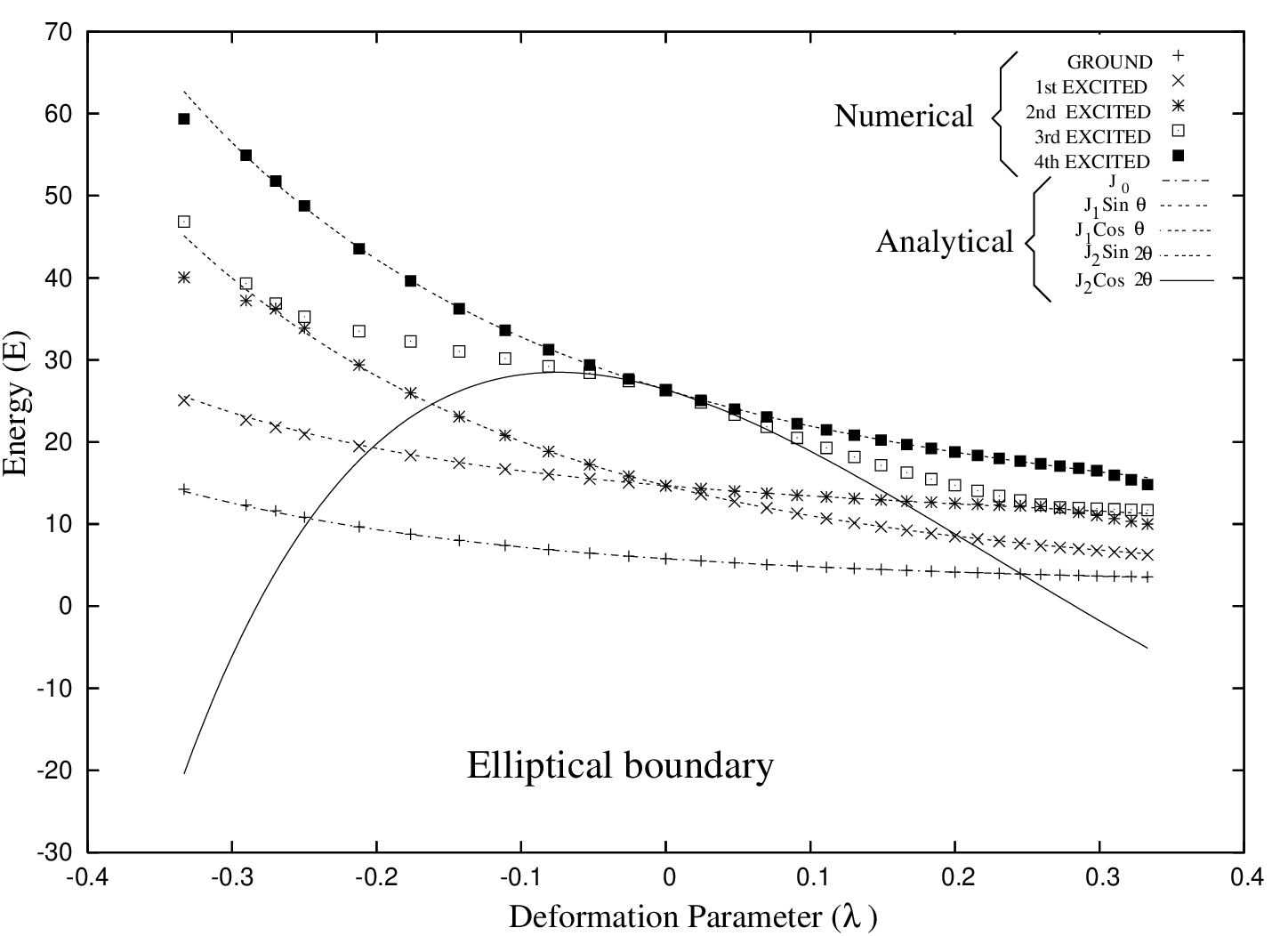}}}
\caption{\label{fig:4} Comparison of the energy eigenvalues
obtained numerically and analytically for a particle confined in
an elliptical boundary shown for the first 5 states (in units of
$\frac{\hbar^{2}}{2mR_{o}^{2}}$). Deformation parameter
$\lambda=\frac{a-b}{a+b}.$}
\end{figure}

 Again $\lambda=0$ describes a circle
with unit radius. Now, $r(\theta)$ in (\ref{eqe1}) can be recast
as,
\begin{equation}
r=R_0[1+\lambda \cos 2\theta -\frac{1}{4}\lambda^2 +
\frac{3}{4}\lambda^2 \cos 4\theta +{\rm
O}(\lambda^3)+...],\label{eqe2}
\end{equation}
with $R_0=\sqrt{ab}$. Comparing with our general Fourier series of
(\ref{gfs}), we observe that,
$$C_2^{(1)}=1,~~
C_{0}^{(2)}=-\frac{1}{4},~~{\rm and}~~C_{4}^{(2)}=\frac{3}{4}.$$
Using (\ref{gse1c}),(\ref{gse2c}),(\ref{ese1cc}),(\ref{ese2cc}),
(\ref{ese1cs}),(\ref{ese2cs}) we find,
\begin{subequations}
\begin{equation}
E_{1}\left(\begin{array}{c}J_{l} \cos l\theta \\ J_{l}\sin l\theta
\end{array}\right)=\left(\begin{array}{c} - \\ +
\end{array}\right)E_{0}\delta_{l1},
\end{equation}
and
$$E_{2}\left(\begin{array}{c}J_{l} \cos l\theta \\ J_{l}\sin l\theta
\end{array}\right)=E_{0}\left(\frac{\delta_{l1}}{2}+1
+\sum_{p,|p-l|=2}\frac{\rho_{l,j}J_{p}'(\rho_{l,j})}{2J_{p}(\rho_{l,j})}\right)$$
\begin{equation}
+\left(\begin{array}{c}
-\frac{3}{4}+\frac{\rho_{l,j}J_{0}'(\rho_{l,j})}
{J_{0}(\rho_{l,j})} \\ \frac{3}{4} \end{array}\right)
E_{0}\delta_{2l},
\end{equation}
\end{subequations}
where $\delta_{ij}$ is Kronecker delta. The results for the
elliptical boundary are shown in FIG.\ref{fig:4}.

 From
FIG.\ref{fig:4} it is seen that as in the case of the supercircle
here also the $J_{1}\cos \theta $ state is strongly affected by
the boundary perturbation and crosses over to its counterpart
$J_{1}\sin \theta$ at $\lambda=0$. However, quite interestingly,
the $J_{2}$ states do not cross but are rather repelled by each
other. They touch each other tangentially at $\lambda=0$. While
for one of these states, $J_2\sin 2\theta$ the analytical method
works quite well, it has a restricted validity for the other one,
viz. $J_2\cos 2\theta$. In fact, we compared the energy levels for
the first 10 states and found out the agreement between the
analytical and the numerical results to be quite satisfactory
except when the levels repel each other. This phenomenon of level
repulsion also goes by the name of ``loci veering" in the
literature. In case of level repulsion the validity of the
perturbation theory for the $J_l\cos l\theta$ states is restricted
to a small range in $\lambda$ (e.g. $|\lambda|\leqq 0.08$). This
is in sharp contrast to the case where there is no repulsion in
which case the agreement with the perturbation
theory persists over a wide range.\\

{\section{Conclusion}}
 One of the principle virtues of the method  proposed is its
 generality. With slight modification, the formalism can readily
 be adopted to study the shape dependence of the eigenvalues of
 a vibrating membrane with Dirichlet conditions on an irregular
 boundary. The approach can also be useful in studying the modes
 of propagation of electromagnetic waves in a waveguide with
 irregular cross section. In fact, recently, Dubertrand et al
 \cite{Dubertrand} have employed a similar scheme for the
 propagation of electromagnetic waves in open dielectric systems.
 Another potential area where this formalism might be useful
 is in the study of quantum dots. This field has been an area
 of vigorous research for the past few years. 2D quantum dots
 are generally taken to have a circular symmetry. However,
 in practice such a symmetry can not be strictly ensured.
 There is bound to be small deviations from exact circular
 symmetry. Hence, probes have been constructed to investigate
 the shape of the dots \cite{Lis, Drouvelis, Prb}. As shown in
 this paper the energy eigenvalues of a particle confined in 2D
 in an infinitely deep potential well will essentially depend
 upon the shape of the confining region. Hence a study of the
 shape dependence of the energy levels might prove to be useful
 in shedding light upon the actual shapes of the dots. Another
 significant aspect of our formalism is the use of the general
 Fourier series to express the deviation of the boundary from
 a circular one which allows us to treat any sort of boundary
 within the limit of small perturbation for which our formalism
 is valid. Even boundaries with sharp singularities can be treated
 in our formalism quite efficiently. For example, 
the square which is a special case of a 
supercircle with $t=1$ can be treated quite
 efficiently by our method. This is borne out by the
accuracy of the results obtained by using our formalism in
 the case of the supercircle for $t=1~(\lambda=1)$,
 which corresponds to 
 a square [FIG.\ref{fig:1} and FIG.\ref{fig:2}]. In fact, to find out the energy 
and the wave function corrections all one needs is to find the Fourier coefficients
 for the closed curve and substitute them in the relevant expressions. Further,
 the corrections to the energy eigenvalue and the eigenfunctions
 are found out exactly in a closed form at each order of
 perturbation without any major approximations which is
 indeed remarkable. The case of the supercircular boundary
 shows that even for quite large perturbations the method
 yields satisfactory results. The accuracy of the method
 can be still improved by including higher order corrections.
 In fact, we have also found out the third order corrections,
 although the results are not included here. On the contrary,
 the case of the elliptical boundary points out to the failure
 of the perturbation theory whenever the energy levels exhibit
 repulsion. This provides potential topics for future
 investigations. Another point which we want to emphasis
 here is that the success (and also the efficiency) of the
 formalism depends to a large extent upon the judicious choice
 of the deformation parameter $\lambda$. For the case of the
 ellipse we defined  $\lambda$ to be equal to $\frac{a-b}{a+b}$
 whereas the eccentricity $\epsilon$ would seem to be a more
 appropriate candidate for $\lambda$. For the elliptical
 boundary we have considered deformations up to the extent
 where $a:b=2:1$ for which $\lambda=0.333$. Had we formulated
 the problem in terms of the eccentricity the same deformation
 would have led to the value of $\epsilon=0.866$. It can also
 be shown that in that case the deformation parameter would
 actually be $\epsilon^{2}$, so that for the same deformation
 we would have $\epsilon^{2}=0.75$ which is obviously much
 larger than the parameter which we have actually used here.
 Such a high value of the deformation parameter goes against
 the very essence of the perturbative nature of the method.
 This means that while we have terminated the Fourier series
 and also the eigenvalues at the second order of smallness
 when working with $\lambda=\frac{a-b}{a+b}$, for
 $\lambda=\epsilon^{2}$ we would have to consider higher
 order terms to get the same accuracy. Finally, we note
 that the same formalism can also be adopted by perturbing
 a square or a rectangular boundary for which the results
 are exactly known.
\begin{acknowledgments}
The authors would like to thank Rahul Sarkar of IIT Kharagpur for
doing the relevant programs for the numerical estimate of the
eigenvalues.
\end{acknowledgments}

\end {document}